\documentclass[12pt,prd,showpacs,tightenlines,nofootinbib]{revtex4}
\usepackage{bm}
\usepackage{graphics}
\usepackage{rotating}
\usepackage{epsfig}

\begin{document}

\title{Relativistic description of weak decays of $B_s$ mesons}
\author{R. N. Faustov}
\author{V. O. Galkin}
\affiliation{Dorodnicyn Computing Centre, Russian Academy of Sciences,
  Vavilov Str. 40, 119333 Moscow, Russia}


\begin{abstract}
The branching fractions of the
semileptonic and rare $B_s$ decays are calculated in the
framework of the QCD-motivated relativistic quark model. The form
factors of the weak $B_s$ transitions are
expressed through the overlap integrals of the initial and final meson
wave functions in the whole accessible kinematical range. The momentum
transfer dependence of the form factors is explicitly determined
without additional model assumptions and extrapolations. The obtained
results agree well with available experimental data.  
\end{abstract}

\maketitle

\section{Introduction}

In recent years significant experimental progress has been achieved
in studying properties of $B_s$ mesons. The Belle Collaboration
considerably increased the number of observed $B_s$ mesons and their
decays due to the data collected in $e^+e^-$ collisions at the
$\Upsilon(10860)$ resonance \cite{belle1}. On the other hand, $B_s$
mesons are copiously produced at Large Hadron Collider
(LHC). First precise data on their properties are coming from the LHCb
Collaboration. Several weak decay modes of the $B_s$
meson were observed for the first time \cite{lhcb1p}.  New data are expected in
near future.

\begin{figure}[b]
  \includegraphics[width=7cm]{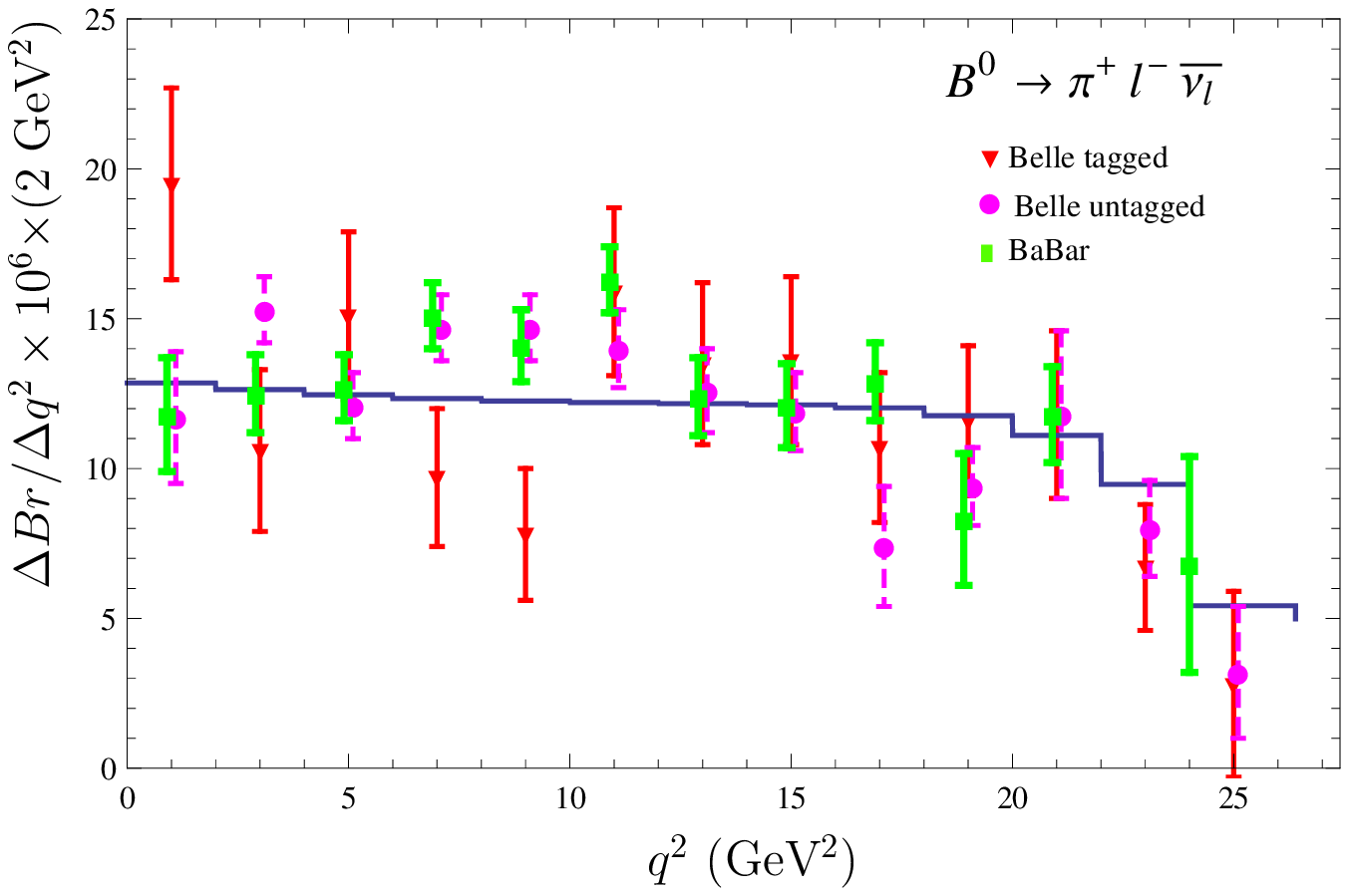} \ \includegraphics[width=7cm]{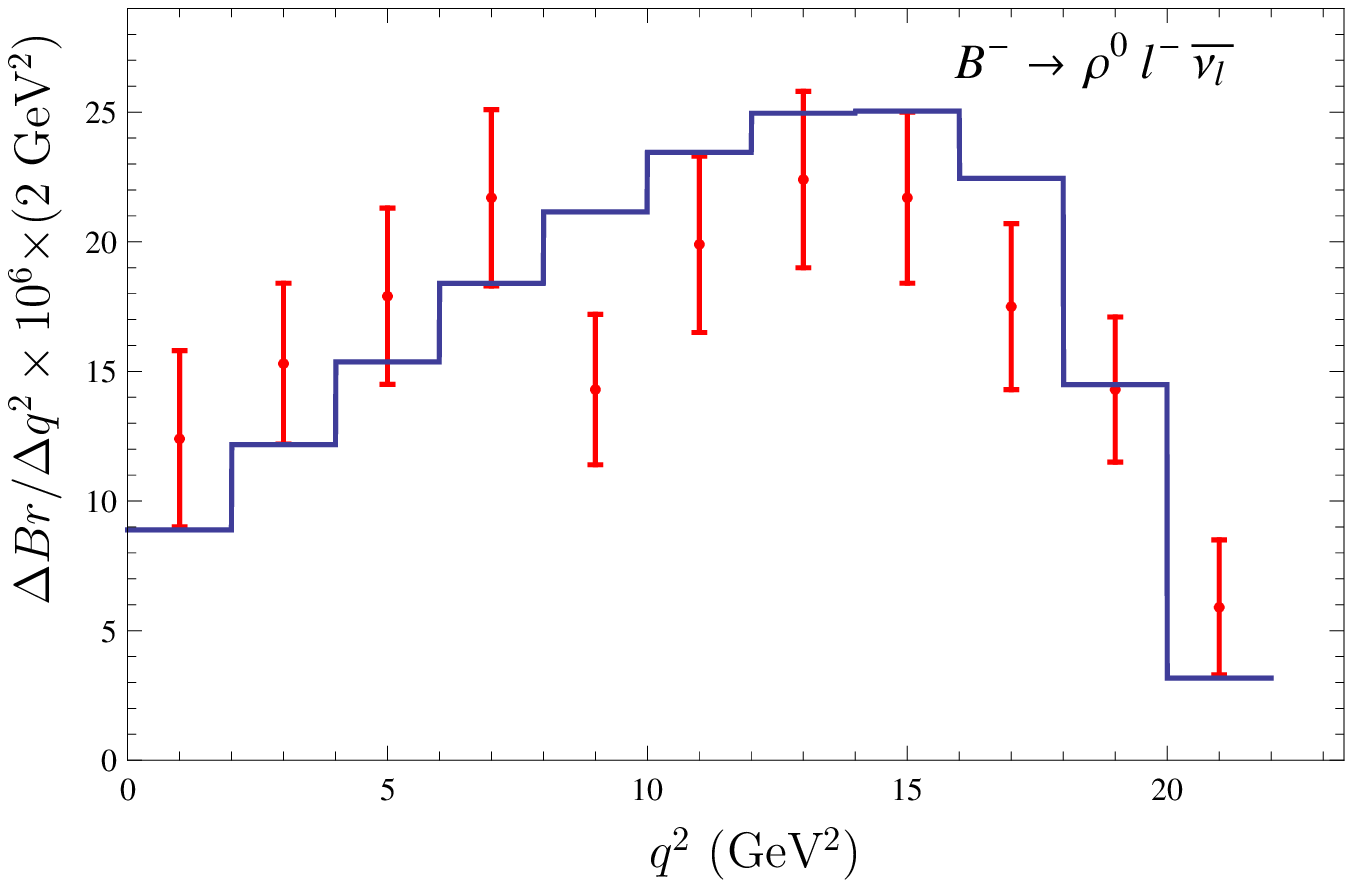}
\caption{Comparison of predictions  of our model with the recent
  experimental data (Belle 2011, 2013; BaBar 2012) for the $B^0\to
  \pi^+ l^-\nu$ decay and Belle (2013) data for the $B\to \rho l\nu$ decay. } 
\label{bpi}
\end{figure}

In this talk we consider the weak $B_s$ transition form factors
and decay rates in
the framework of the relativistic quark model based on the
quasipotential approach and quantum chromoynamics (QCD) \cite{fg2013}. We previously
applied this model for the calculation of the weak $B$ transitions
\cite{hlsem}. Recently Belle  and BaBar 
Collaborations \cite{expbpi} published new more precise data on differential
distributions in $B\to\pi l\nu_l$ and $B\to\rho l\nu_l$ decays. In
Fig.~\ref{bpi} we compare predictions of our model with these data.
From this figure we see that our predictions agree well with new
data. The fit  of our model predictions to the combined Belle and
BaBar data yields the following values of the CKM matrix element $V_{ub}$ 
\begin{itemize}
\item $B\to \pi l\nu_l$ decays
$ \qquad\qquad\qquad\quad\
  |V_{ub}|=(4.07\pm0.07_{\rm exp}\pm0.21_{\rm theor})\times 10^{-3}
$
\item $B\to \rho l\nu_l$ decays
$\qquad\qquad\qquad\quad\
  |V_{ub}|=(4.03\pm0.15_{\rm exp}\pm0.21_{\rm theor})\times 10^{-3}
$
\item combined data on $B\to \pi(\rho) l\nu_l$ 
$\quad
{  |V_{ub}|=(4.06\pm0.06_{\rm exp}\pm0.21_{\rm theor})\times 10^{-3}}
$
\end{itemize}
These values are in good agreement with the averaged value extracted
from the inclusive $B$
decays \cite {pdg}
 $\quad
{|V_{ub}|=(4.41\pm0.15^{+0.15}_{-0.19})\times 10^{-3}}. 
$

\section{Relativistic quark model}

All considerations in this talk are done in the framework of the
relativistic quark model.  
The model is based on the
quasipotential approach in quantum field theory with the QCD motivated
interaction.  Hadrons are considered as the bound states of
constituent quarks and are described by the
single-time wave functions satisfying the
three-dimensional Schr\"odinger-like equation, which is
relativistically invariant \cite{mass}: 
\begin{equation}
\label{quas}
{\left(\frac{b^2(M)}{2\mu_{R}}-\frac{{\bf
p}^2}{2\mu_{R}}\right)\Psi_{M}({\bf p})} =\int\frac{d^3 q}{(2\pi)^3}
 V({\bf p,q};M)\Psi_{M}({\bf q}),
\end{equation}
where
\begin{equation}
\mu_{R}=\frac{M^4-(m^2_1-m^2_2)^2}{4M^3},\qquad {b^2(M) }
=\frac{[M^2-(m_1+m_2)^2][M^2-(m_1-m_2)^2]}{4M^2},
\end{equation}
 $M$ is the meson mass, $m_{1,2}$ are the quark masses,
and ${\bf p}$ is their relative momentum.  
The interaction quasipotential $V({\bf p,q};M)$ consists of the perturbative one-gluon
exchange part and the nonperturbative confining part \cite{mass}. The Lorentz
structure of the latter part includes the scalar and vector linearly
rising interactions. The vertex of the vector interaction contains the Pauli
term (the nonperturbative anomalous chromomagnetic moment of the quark) which
enables vanishing of the spin-dependent chromomagnetic interaction in
accord with the flux tube model. 

For the consideration of meson weak decays it is necessary to
calculate the matrix element of the weak current between meson
states. In the quasipotential approach such a matrix element between a $B_s$ meson with mass $M_{B_s}$ and
momentum $p_{B_s}$ and a final $F$ meson with mass $M_{F}$ and
momentum $p_{F}$ is given by  \cite{mass}
\begin{equation}\label{mxet} 
\langle F(p_{F}) \vert J^W_\mu \vert B_s(p_{B_s})\rangle
=\int \frac{d^3p\, d^3q}{(2\pi )^6} \bar \Psi_{{F}\,{\bf p}_F}({\bf
p})\Gamma _\mu ({\bf p},{\bf q})\Psi_{B_s\,{\bf p}_{B_s}}({\bf q}),
\end{equation}
where $\Gamma _\mu ({\bf p},{\bf
q})$ is the two-particle vertex function and  
$\Psi_{M\,{\bf p}_M}({\bf p})$ are the
meson ($M=B_s,{F})$ wave functions projected onto the positive energy states of
quarks and boosted to the moving reference frame with the total momentum ${\bf
  p}_M$, and  ${\bf p},{\bf q}$ are relative quark momenta. 

The explicit expression for the vertex function  $\Gamma_\mu ({\bf
  p},{\bf q})$ can be found in Ref.~\cite{fg2013}. It contains
contributions both from the leading order spectator diagram and from
subleading order diagrams accounting for the contributions of the
negative-energy intermediate states. The leading order contribution
contains the $\delta$ function which allows us to take one of the
integrals in the matrix element (\ref{mxet}).  Calculation of the subleading order
contribution is more complicated due to the dependence on
the relative momentum in the energies of the initial heavy and final
light quarks. For the energy of the heavy quarks we use the heavy quark
expansion. For the light quark such expansion is not applicable.  However, 
if the final $F$ meson is light ($K$, $\varphi$ etc.) than it has a large (compared to its
mass) recoil momentum 
($|{\bm\Delta}_{\rm  max}|=(M_{B_s}^2-M_F^2)/(2M_{B_s})\sim 2.6$~GeV) in  almost
the whole kinematical range except the small region near  
$q^2=q^2_{\rm max}$ ($|{\bm\Delta}|=0$).  This also means that the
recoil momentum of the final meson is large with respect to the mean
relative quark momentum $|{\bf p}|$ in the meson  ($\sim 0.5$~GeV).
Thus one can neglect  $|{\bf p}|$ compared to $|{\bm\Delta}|$ in  the
light quark energies
$\epsilon_{q}(p+\Delta)=\sqrt{m_{q}^2+({\bf 
p}+{\bm\Delta})^2}$, replacing it with  $\epsilon_{q}(\Delta)=
\sqrt{m_{q}^2+{\bm\Delta}^2}$  in expressions for the
subleading contribution.  Such replacement removes the relative
momentum dependence of the quark energies  and thus permits the
performance of one of the integrations in the subleading 
contribution using the quasipotential equation. Since the subleading
contributions are suppressed the uncertainty introduced by such
procedure is small.  As a result, the weak
decay matrix element is expressed through the usual overlap integral
of initial and final meson wave functions and its momentum dependence
can be determined in the whole accessible kinematical range without
additional assumptions. 

\section{Semileptonic $B_s$ decays to $D_s$ mesons}
\label{sec:ffsdgs}

The matrix elements of weak current $J^W$ between meson ground states are
usually parametrized by the following set of the invariant form factors
\begin{equation}
  \label{eq:pff1}
  \langle D_s(p_{D_s})|\bar c \gamma^\mu b|B_s(p_{B_s})\rangle
  =f_+(q^2)\left[p_{B_s}^\mu+ p_{D_s}^\mu-
\frac{M_{B_s}^2-M_{D_s}^2}{q^2}\ q^\mu\right]+
  f_0(q^2)\frac{M_{B_s}^2-M_{D_s}^2}{q^2}\ q^\mu,
\end{equation}
\begin{equation}
\label{eq:pff2} 
 \langle D_s(p_{D_s})|\bar c \gamma^\mu\gamma_5 b|B_s(p_{B_s})\rangle
  =0,
\end{equation}
\begin{eqnarray}
  \label{eq:vff1}
  \langle {D^*_s}(p_{D^*_s})|\bar c \gamma^\mu b|B(p_{B_s})\rangle\!\!\!\!&=\!\!\!\!
  &\frac{2iV(q^2)}{M_{B_s}+M_{D^*_s}} \epsilon^{\mu\nu\rho\sigma}\epsilon^*_\nu
  p_{B_s\rho} p_{{D^*_s}\sigma},\\ 
\label{eq:vff2}
\langle {D^*_s}(p_{D^*_s})|\bar c \gamma^\mu\gamma_5 b|B_s(p_{B_s})\rangle\!\!\!\!&=\!\!\!\!&2M_{D^*_s}
A_0(q^2)\frac{\epsilon^*\cdot q}{q^2}\ q^\mu
 +(M_{B_s}+M_{D^*_s})A_1(q^2)\left(\epsilon^{*\mu}-\frac{\epsilon^*\cdot
    q}{q^2}\ q^\mu\right)\cr
&&-A_2(q^2)\frac{\epsilon^*\cdot q}{M_{B_s}+M_{D^*_s}}\left[p_{B_s}^\mu+
  p_{D^*_s}^\mu-\frac{M_{B_s}^2-M_{D^*_s}^2}{q^2}\ q^\mu\right]. 
\end{eqnarray}
At the maximum recoil point ($q^2=0$) these form
factors satisfy the following conditions: 
\[f_+(0)=f_0(0),\qquad
A_0(0)=\frac{M_{B_s}+M_{D^*_s}}{2M_{D^*_s}}A_1(0)
-\frac{M_{B_s}-M_{D^*_s}}{2M_{D^*_s}}A_2(0).\]

To calculate the weak decay matrix element  we employ the
heavy quark and large recoil expansion, which permits us to take one of the integrals in
the subleading contribution of the vertex function  to the weak
current matrix element. As a result we express all 
matrix elements through the usual overlap integrals of the meson wave
functions. We find that the decay form factors can
be approximated with sufficient accuracy by the following expressions:
\begin{eqnarray}
  \label{fitfv}
&& f_+(q^2),\ V(q^2),\ A_0(q^2)=F(q^2) 
=\frac{F(0)}{\left(1-\frac{q^2}{ M^2}\right)
    \left(1-\sigma_1 
      \frac{q^2}{M_{B_c^*}^2}+ \sigma_2\frac{q^4}{M_{B_c^*}^4}\right)},\\
  \label{fita12}
&& f_0(q^2),\ A_1(q^2),\ A_2(q^2)=F(q^2)
=\frac{F(0)}{ \left(1-\sigma_1
      \frac{q^2}{M_{B_c^*}^2}+ \sigma_2\frac{q^4}{M_{B_c^*}^4}\right)},
\end{eqnarray}
where $M=M_{B_c^*}=6.332$~GeV  for the form factors $f_+(q^2),V(q^2)$ and
$M=M_{B_c}=6.272$~GeV for the form factor $A_0(q^2)$; the values  $F(0)$ and $\sigma_{1,2}$ are given in 
Table~\ref{hhff}. 
\begin{table}
\caption{Form factors of weak $B_s\to D_s^{(*)}$ transitions. }
\label{hhff}
\begin{tabular}{ccccccc}
\hline
   &\multicolumn{2}{c}{{$B_s\to D_s$}}&\multicolumn{4}{c}{{\  $B_s\to D^*_s$\
     }}\\
\cline{2-3}  \cline{4-7}
& $f_+$ & $f_0$& $V$ & $A_0$ &$A_1$&$A_2$ \\
\hline
$F(0)$          &0.74&0.74 &  0.95 & 0.67 & 0.70& 0.75\\
$F(q^2_{\rm max})$&1.15&0.88 &  1.50 & 1.06& 0.84& 1.04 \\
$\sigma_1$      &0.200&0.430& 0.372 &0.350&  0.463& 1.04\\
$\sigma_2$
&$-0.461$&$-0.464$&$-0.561$&$-0.600$&$-0.510$&$-0.070$\\
\hline
\end{tabular}

\end{table}
The values
of $\sigma_{1,2}$ are determined with a few tenths of percent
errors. The main uncertainties of the form factors originate from 
the account of $1/m_Q^2$ corrections at zero recoil only and from
the higher order $1/m_Q^3$ contributions and can be roughly
estimated in our approach to be about 2\%.

In Table \ref{compbpiff} we confront our predictions for the form factors of 
semileptonic decays $B_s\to  D_s^{(*)} e\nu$  at maximum recoil point
$q^2=0$ with results of other approaches
\cite{kp,bcnp,cfkw,llw,lsw}.
\begin{table}
\caption{Comparison of theoretical predictions for the form factors of 
  semileptonic decays $B_s\to  D_s^{(*)} e\nu$  at maximum
  recoil point $q^2=0$.  }
\label{compbpiff}
\begin{tabular}{cccccc}
\hline
     & $f_+(0)$ & $V(0)$ & $A_0(0)$ &$A_1(0)$&$A_2(0)$ \\
\hline
our      &$0.74\pm0.02$  & $0.95\pm0.02$  & $0.67\pm0.01$  & $0.70\pm0.01$ & $0.75\pm0.02$\\
\cite{kp}&0.61  & 0.64  &       & 0.56 & 0.59\\
\cite{bcnp} &$0.7\pm0.1$ &$0.63\pm0.05$ &$0.52\pm0.06$ &$0.62\pm0.01$&$0.75\pm0.07$ \\
\cite{cfkw} &$0.57^{+0.02}_{-0.03}$ &$0.70^{+0.05}_{-0.04}$ &
&$0.65^{+0.01}_{-0.01}$&$0.67^{+0.01}_{-0.01}$  \\
\cite{llw} & $0.86^{+0.17}_{-0.15}$&&&&\\
\cite{lsw} & &$0.74^{+0.05}_{-0.05}$ &$0.63^{+0.04}_{-0.04}$ 
&$0.61^{+0.04}_{-0.04}$&$0.59^{+0.04}_{-0.04}$\\
\hline
\end{tabular}
\end{table}
Different quark models are used in
Refs.~\cite{kp,cfkw,lsw}, while the QCD and light cone sum rules are
employed in Refs.~\cite{bcnp,llw}. We find that these significantly
different theoretical calculations lead to rather similar
decay form factors. One of the main advantages of our model is its
ability not only to obtain the decay form factors at a single
kinematical point, but also to determine its $q^2$ dependence in the
whole range without any additional assumptions or extrapolations.

Using these weak decay form factors we calculate the total semileptonic decay
rates. It is necessary to point out that the kinematical
range  accessible in these semileptonic decays is rather
broad. Therefore the knowledge of the $q^2$ dependence of the form
factors is very important for reducing theoretical uncertainties of the decay
rates. Our results for the semileptonic $B_s\to D_s^{(*)} l\nu$ decay
rates are given in Table~\ref{comphlff} in comparison 
with previous calculations. The authors of Ref.\cite{bcnp} use the QCD
sum rules, while the light cone sum rules approach is adopted in
Ref.~\cite{llw}. Different types of constituent quark models are
employed in Refs.~\cite{lsw,cfkw,zll} and the three point QCD sum
rules are used in Ref.~\cite{ab}. We see that our predictions are
consistent with results of quark model calculations in
Refs.~\cite{lsw,cfkw}. They are almost two times larger than
the QCD sum rules and light cone sum rules results of
Refs.~\cite{bcnp,llw}, but slightly lower than the values of
Refs.~\cite{zll,ab}.
\begin{table}
\caption{Comparison of theoretical predictions for the branching fractions of semileptonic
  decays $B_s\to D_s^{(*)} l\nu$ (in \%).  }
\label{comphlff}
\begin{tabular}{cccccccc}
\hline
Decay& this paper & \cite{bcnp}&\cite{cfkw}&\cite{llw}  &\cite{lsw} &\cite{zll}
 & \cite{ab}\\
\hline
$B_s\to D_se\nu$& $2.1\pm0.2$ & $1.35\pm0.21$ & 1.4-1.7&$1.0^{+0.4}_{-0.3}$&  &
2.73-3.00& 2.8-3.8\\ 
$B_s\to D_s\tau\nu$& $0.62\pm0.05$ & & 0.47-0.55 &$0.33^{+0.14}_{-0.11}$&  &&\\
$B_s\to D_s^*e\nu$& $5.3\pm0.5$ & $2.5\pm0.1$& 5.1-5.8 && $5.2\pm0.6$  &
7.49-7.66& 1.89-6.61\\
$B_s\to D_s^*\tau\nu$& $1.3\pm0.1$ &  & 1.2-1.3&&$1.3^{+0.2}_{-0.1}$&
&\\
\hline
\end{tabular}

\end{table} 
We find that the total branching fraction of the semileptonic decays of
$B_s$ mesons to the ground state $D_s^{(*)}$ is equal to
$Br(B_s\to D_s^{(*)} e\nu)=(7.4\pm 0.7)\%$ and $Br(B_s\to D_s^{(*)} \tau\nu)=(1.92\pm 0.15)\%$.

Using the same approach we calculate the form factors of $B_s$ decays
to radially and orbitally excited $D_s$ mesons.  The predictions
for the branching fractions of $B_s$ decays
to radially excited $D_s$ mesons are given in
Table~\ref{compreff}.
 We find that the decay rates of the semileptonic $B_s$ decays to the
pseudoscalar $D_s(2S)$ and vector $D_s^*(2S)$ mesons have close
values. The
total contribution of these decays is obtained to be $Br(B_s\to
D_s^{(*)}(2S)e\nu)=(0.65\pm0.06)\%$ and $Br(B_s\to
D_s^{(*)}(2S)\tau\nu)=(0.026\pm0.003)\%$.
\begin{table}
\caption{Predictions for the branching fractions of semileptonic
  decays $B_s\to D_s^{(*)}(2S) l\nu$ (in \%).  }
\label{compreff}
\begin{tabular}{cc}
\hline
Decay& Br \\
\hline
$B_s\to D_s(2S)e\nu$& $0.27\pm0.03$ \\ 
$B_s\to D_s(2S)\tau\nu$& $0.011\pm0.001$\\
$B_s\to D_s^*(2S)e\nu$& $0.38\pm0.04$\\
$B_s\to D_s^*(2S)\tau\nu$& $0.015\pm0.002$ \\
\hline
\end{tabular}

\end{table}

Our
predictions for the branching fractions of the semileptonic $B_s$
decays to orbitally excited $D_s$ mesons are
given in Table~\ref{comphlffoe} in comparison with other
calculations. We find that decays to
$D_{s1}$ and $D^*_{s2}$ mesons are dominant. First we compare with our previous calculation
\cite{orbexc} which was performed in the framework of the heavy quark
expansion. We present results found in the infinitely heavy quark limit
($m_Q\to\infty$) and with the account of first order $1/m_Q$
corrections. It was argued  \cite{orbexc} that $1/m_Q$ corrections are large and
their inclusion significantly influences the decays rates. The large
effect of subleading heavy quark corrections was found to be a
consequence of the vanishing of the leading 
order contributions to the decay matrix elements, due to heavy quark
spin-flavour symmetry, at the point of zero recoil of the final charmed
meson, while the subleading order contributions  do not vanish at this
kinematical point. Here we calculated the decay rates without
application of the heavy quark expansion. We find that 
nonperturbative results agree well with the ones
obtained with the account of the leading order $1/m_Q$ corrections \cite{orbexc}. This
means that the  higher order in $1/m_Q$ corrections are small, as was
expected. Then we compare our predictions with the results of
calculations within other approaches. The authors of
Refs.~\cite{saefhp,zll} employ different types of constituent
quark models for their calculations. Light cone and three point QCD sum rules
are used in Refs.~\cite{llw}. In general we find reasonable
agreement between our predictions and results of
Refs.~\cite{saefhp,llw}, but results of the quark
model calculations \cite{zll} are slightly larger. The total semileptonic
decay branching fractions to orbitally excited $D_{s}$ mesons are found
to be  $Br(B_s\to D_{sJ}^{(*)}e\nu)= (2.1\pm0.2)\%$ and $Br(B_s\to
D_{sJ}^{(*)}\tau\nu)= (0.11\pm0.01)\%$.

The first experimental measurement of the semileptonic decay $B_s\to
D_{s1}\mu\nu$ was done by the D0 Collaboration \cite{d0}. The
branching fraction was obtained by assuming that the $D_{s1}$ 
production in semileptonic decay comes entirely from the $B_s$ decay  and using
a prediction for $Br(D_{s1}\to D^*K^0_S)=0.25$. Its value $Br(B_s \to
D_{s1} X\mu\nu)_{\rm D0}=(1.03\pm 0.20\pm 0.17\pm0.14)\%$ is in good agreement
with our prediction $0.84\pm0.09$ given in Table~\ref{comphlffoe}.

\begin{table}
\caption{Comparison of the predictions for the branching fractions of the semileptonic
  decays $B_s\to D_{sJ}^{(*)} l\nu$ (in \%).  }
\label{comphlffoe}
\hspace*{-0.1cm}\begin{tabular}{ccccccc}
\hline
Decay& this paper& $m\to\infty$ & with
  & \cite{saefhp} &\cite{zll}&\cite{llw}\\
&&\cite{orbexc}&$1/m_Q$ \cite{orbexc}&&&\\
\hline
$B_s\to D_{s0}^*e\nu$&      $0.36\pm0.04$& 0.10 & 0.37 & 0.443 & 0.49-0.571&
$0.23^{+0.12}_{-0.10}$\\ 
$B_s\to D_{s0}^*\tau\nu$&  $0.019\pm0.002$&   &  &  & & $0.057^{+0.028}_{-0.023}$
\\ 
$B_s\to D_{s1}'e\nu$&      $0.19\pm0.02$& 0.13 & 0.18 & 0.174-0.570 &
0.752-0.869&   \\
 $B_s\to D_{s1}'\tau\nu$&      $0.015\pm0.002$ &&&&& \\ 
$B_s\to D_{s1}e\nu$&      $0.84\pm0.09$& 0.36 & 1.06 & 0.477 &&  \\
 $B_s\to D_{s1}\tau\nu$&    $0.049\pm0.005$&  &&&& \\
 $B_s\to D_{s2}^*e\nu$&      $0.67\pm0.07$& 0.56 & 0.75 & 0.376 & & \\ 
 $B_s\to D_{s2}^*\tau\nu$&      $0.029\pm0.003$  &&&&& \\ 
\hline
\end{tabular}

\end{table} 

Recently the LHCb Collaboration \cite{lhcb3} reported the first 
observation of the orbitally excited $D_{s2}^*$ meson in the
semileptonic $B_s$ decays. The decay to the $D_{s1}$ meson was also
observed. The measured branching fractions relative 
to the total $B_s$ semileptonic rate are $Br(B_s\to
D_{s2}^*X\mu\nu)/Br(B_s\to X\mu\nu)_{\rm LHCb} =(3.3\pm1.0\pm0.4)\%,$   $Br(B_s\to
D_{s1}X\mu\nu)/Br(B_s\to X\mu\nu)_{\rm LHCb} =(5.4\pm1.2\pm0.5)\%.$ The
$D_{s2}^*/D_{s1}$ event ratio is found to be $Br(B_s\to
D_{s2}^*X\mu\nu)/Br(B_s\to D_{s1}X\mu\nu)_{\rm LHCb}=0.61\pm0.14\pm0.05.$ These
values can be compared with our predictions if we assume that decays
to $D_{s1}$ and $D_{s2}^*$ mesons give dominant contributions to the ratios. Summing up the semileptonic
$B_s$ decay branching fractions to the ground state, first radial and
orbital excitations of $D_s$ mesons we get for the total $B_s$ semileptonic rate
$Br(B_s\to X\mu\nu)=(10.2\pm 1.0)\%$. Then using the calculated values
from Table~\ref{comphlffoe} we get  $Br(B_s\to
D_{s2}^*\mu\nu)/Br(B_s\to X\mu\nu)_{\rm theor} =(6.5\pm1.2)\%,$  $Br(B_s\to
D_{s1}\mu\nu)/Br(B_s\to X\mu\nu)_{\rm theor} =(8.2\pm1.6)\%,$ and $Br(B_s\to
D_{s2}^*\mu\nu)/Br(B_s\to D_{s1}\mu\nu)_{\rm theor}=0.79\pm0.14.$ 
The predicted central values are larger than 
experimental ones, but the results agree with experiment within $2\sigma$.  

The following total semileptonic $B_s$ branching
fractions were found: 
(1)  for decays to ground state $D_s^{(*)}$ mesons  $Br(B_s\to
D_s^{(*)} e\nu)=(7.4\pm 0.7)\%$ and $Br(B_s\to D_s^{(*)}
\tau\nu)=(1.92\pm 0.15)\%$;
(2) for decays to radially excited
$D_s^{(*)}(2S)$ mesons  $Br(B_s\to
D_s^{(*)}(2S)e\nu)=(0.65\pm0.06)\%$ and $Br(B_s\to
D_s^{(*)}(2S)\tau\nu)=(0.026\pm0.003)\%$;
(3) for decays to orbitally excited $D_{sJ}^{(*)}$ mesons 
$Br(B_s\to D_{sJ}^{(*)}e\nu)= (2.1\pm0.2)\%$ and $Br(B_s\to
D_{sJ}^{(*)}\tau\nu)= (0.11\pm0.01)\%$.
 We see that these
branching fractions significantly decrease with excitation. Therefore,
we can conclude that considered decays give the dominant contribution to the
total semileptonic branching fraction $Br(B_s\to D_s e\nu+{\rm
  anything})$. Summing up these contributions we get the value
$(10.2\pm 1.0)\%$, which agrees with the experimental value of   
$Br(B_s\to D_s e\nu+{\rm anything})_{\rm Exp.}=(7.9\pm2.4)\%$
\cite{pdg}.

\section{Charmless semileptonic $B_s$ decays}
\label{sdbsk}
\begin{table}
\caption{Calculated form factors of weak $B_s\to K^{(*)}$ transitions.
}
\label{hffl}
\begin{tabular}{ccccccccccc}
\hline
   &\multicolumn{3}{c}{{$B_s\to K$}}&\multicolumn{7}{c}{{\  $B_s\to K^*$\
     }}\\
\cline{2-4} \cline{5-11}
& $f_+$ & $f_0$& $f_T$& $V$ & $A_0$ &$A_1$&$A_2$& $T_1$ & $T_2$& $T_3$\\
\hline
$F(0)$          &0.284 &0.284 &  0.236 & 0.291 & 0.289& 0.287 & 0.286 & 0.238& 0.238& 0.122\\
$F(q^2_{\rm max})$&5.42  &0.459 &  0.993 & 3.06& 2.10& 0.581 &  0.953 & 1.28& 0.570& 0.362\\
$\sigma_1$      &$-0.370$&$-0.072$& $-0.442$& $-0.516$ &$-0.383$&  0&
1.05  &$-1.20$&$0.241$& $0.521$\\
$\sigma_2$      &$-1.41$&$-0.651$&$0.082$&$-2.10$&$-1.58$&$-1.06$&
0.074  &$-2.44$&$-0.857$& $-0.613$\\
\hline
\end{tabular}

\end{table}

Comparing the invariant form factor decomposition (\ref{eq:pff1})--(\ref{eq:vff2})  with the results of the calculations of
the weak current matrix element in our model we determine the form
factors in the whole accessible kinematical range through the overlap
integrals of the meson wave functions. The explicit expressions are
given in Ref.~\cite{fg2013}. For the numerical evaluations of
the corresponding overlap integrals we use the quasipotential wave
functions of $B_s$ and $K^{(*)}$ mesons obtained in their mass spectra
calculations \cite{mass}. The weak $B_s\to K^{(*)}$ transition form
factors can be approximated with good accuracy by
Eqs.~(\ref{fitfv}), (\ref{fita12}).  The obtained values $F(0)$ and
$\sigma_{1,2}$ are given in Table~\ref{hffl}.

\begin{table}
\caption{Comparison of theoretical predictions for the branching
  fractions of semileptonic decays $B_s\to K^{(*)} l\nu_l$ (in $10^{-4}$).  }
\label{compslbsk}
\begin{tabular}{cccc}
\hline
Decay& this paper & \cite{wx}&\cite{wzz} \\
\hline
$B_s\to K e\nu_e$& $1.64\pm0.17$ & $1.27^{+0.49}_{-0.30}$  &$1.47\pm0.15$ \\ 
$B_s\to K\tau\nu_\tau$& $0.96\pm0.10$ &$0.778^{+0.268}_{-0.201}$   &$1.02\pm0.11$\\
$B_s\to K^*e\nu_e$& $3.47\pm0.35$ & & $2.91\pm0.26$ \\
$B_s\to K^*\tau\nu_\tau$& $1.67\pm0.17$ & &$1.58\pm0.13$\\
\hline
\end{tabular}

\end{table}
Using these form factors  we get predictions for the  total
decay rates.  The kinematical range accessible in the heavy-to-light
$B_s\to K^{(*)}$ transitions is very 
broad, making knowledge of the $q^2$ dependence of the
form factors to be an important issue. Therefore, the explicit determination of
the momentum dependence of the weak decay form factors in the whole
$q^2$ range without any additional assumptions is an important
advantage of our model.  
The 
calculated branching fractions of the semileptonic $B_s\to
K^{(*)}l\nu_l$ decays are presented in Table~\ref{compslbsk} in
comparison  with other theoretical predictions \cite{wx,wzz}. The
perturbative QCD factorization approach is used in Ref.~\cite{wx},
while in Ref.~\cite{wzz} light cone sum rules are employed. From the
comparison in Table~\ref{compslbsk} we see that all theoretical
predictions for the $B_s$ semileptonic branching fractions agree
within uncertainties. This is not surprising since these significantly
different approaches predict close behavior of the corresponding weak
form factors.  
 
We employ the same approach for the calculation of the form factors of
the weak $B_s$ decays to orbitally excited $K_{J}^{(*)}$ mesons. The total semileptonic  $B_s\to K_J^{(*)} l\nu_l$ branching
fractions are given in Table~\ref{comphlffk}. We see that our
model predicts close values (about $1\times10^{-4}$) for all
semileptonic $B_s$ branching fractions to the first orbitally excited $K_{J}^{(*)}$
mesons. Indeed, the difference between branching fractions is less than
a factor of 2. This result is in contradiction to the dominance of specific
modes (by more than a factor of 4) in the heavy-to-heavy semileptonic $B\to D_J^{(*)}
l\nu_l$ and  $B_s\to D_{sJ}^{(*)} l\nu_l$ decays, but it is consistent with predictions for
the corresponding
heavy-to-light semileptonic $B$ decays to orbitally excited light
mesons \cite{borbldecay}. The above mentioned suppression of some
heavy-to-heavy decay channels to orbitally excited heavy mesons was
mostly pronounced in the heavy quark limit and then slightly reduced by the
heavy quark mass corrections which are found to be large. Thus our result once again indicates 
that the $s$ quark cannot be treated as a heavy one and should be
considered to be light instead, as we always did in our calculations.

In Table~\ref{comphlffk} we compare our predictions for the
semileptonic $B_s$ branching fractions to orbitally excited $
K_J^{(*)}$ mesons with previous calculations
\cite{ymz,wal,llww,ykc,llw,w}. The consideration in
Ref.~\cite{ymz} is based on QCD sum rules. The light cone sum rules are
used in Refs.~\cite{wal,ykc}, while Refs.~\cite{llww,llw,w} employ the
perturbative QCD approach. Reasonable agreement between our results
and other predictions \cite{ymz,wal,w} is
observed for the semileptonic $B_s$ decays to the scalar and tensor $K$
mesons. The values  of Ref.~\cite{llww} are almost a factor 3
higher. For the semileptonic $B_s$ decays to axial vector $K$
mesons predictions are significantly different even within rather
large errors. Therefore experimental measurement of these decay
branching fractions can help to discriminate between theoretical
approaches.
\begin{table}
\caption{Comparison of theoretical predictions for the branching
  fractions of semileptonic decays  $B_s\to K_J^{(*)} l\nu_l$ (in $10^{-4}$).  }
\label{comphlffk}
\begin{tabular}{cccccccc}
\hline
Decay& this paper &\cite{ymz}& \cite{wal}&\cite{llww}  &\cite{ykc} &\cite{llw}&\cite{w}
 \\
\hline
$B_s\to K^*_0 e\nu_e$& $0.71\pm0.14$&  $0.36^{+0.38}_{-0.24}$& $1.3^{+1.3}_{-0.4}$& $2.45^{+1.77}_{-1.05}$ &  
& &  \\
$B_s\to K^*_0 \tau\nu_\tau$& $0.21\pm0.04$&& $0.52^{+0.57}_{-0.18}$& $1.09^{+0.82}_{-0.47}$   &  
& &  \\
$B_s\to K_1(1270) e\nu_e$& $1.41\pm0.28$  & & & &  $4.53^{+1.67}_{-2.05}$&
$5.75^{+3.49}_{-2.89}$&\\
$B_s\to K_1(1270) \tau\nu_\tau$& $0.30\pm0.06$ & &&& & $2.62^{+1.58}_{-1.31}$
&\\
$B_s\to K_1(1400) e\nu_e$& $0.97\pm0.20$ & & &  &$3.86^{+1.43}_{-1.75}$&
$0.03^{+0.05}_{-0.02}$& \\
$B_s\to K_1(1400) \tau\nu_\tau$& $0.25\pm0.05$ & & & && $0.01^{+0.02}_{-0.01}$
&\\
$B_s\to K^*_2 e\nu_e$& $1.33\pm0.27$ & & & & &&$0.73^{+0.48}_{-0.33}$  \\
$B_s\to K^*_2 \tau\nu_\tau$& $0.36\pm0.07$ & & & &&&
$0.25^{+0.17}_{-0.12}$ \\
\hline
 \end{tabular}

\end{table}

We see that total branching fractions of semileptonic
$B_s$ decays to the ground state and first orbitally excited $K$ mesons have
close values of about $5\times 10^{-4}$. Summing up these contributions,
we get $(9.5\pm1.0)\times 10^{-4}$. This value is almost 2 orders of magnitude
lower than our prediction for the corresponding sum of branching
fractions of the semileptonic $B_s$ to $D_s$
mesons  as it was expected from the ratio of CKM matrix
elements $|V_{ub}|$ and $|V_{cb}|$. Therefore the total semileptonic $B_s$ decay
branching fraction is dominated by the decays to $D_s$ mesons and in our model is equal 
to $(10.3\pm1.0)\%$  in agreement with the experimental value of
$Br(B_s\to X e\nu_e)_{\rm Exp.}=(9.5\pm2.7)\%$~\cite{pdg}. 

\section{Rare semileptonic $B_s$ decays}
\begin{figure}
 \includegraphics[width=6.8cm]{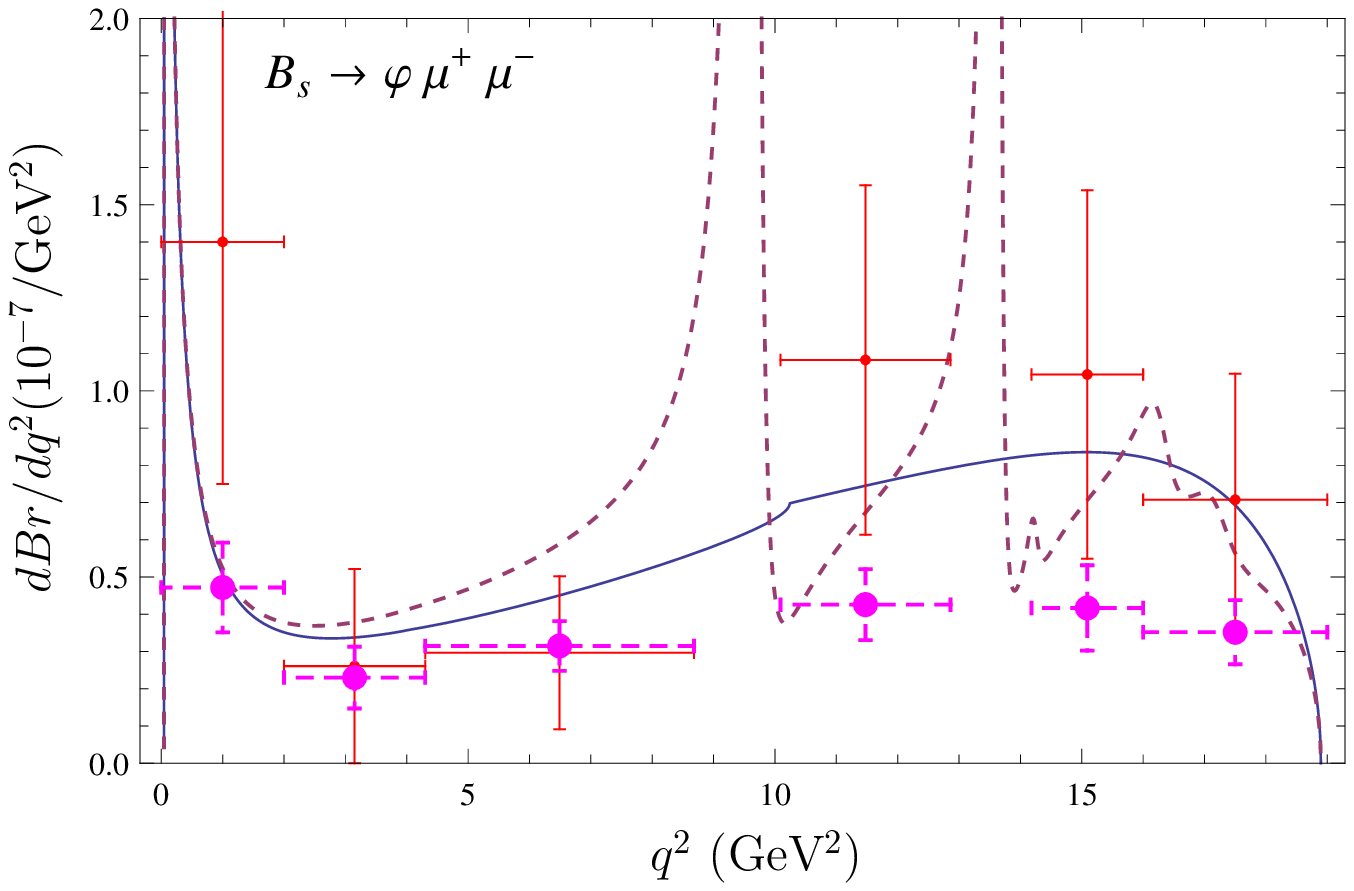} \ \ \  \includegraphics[width=6.8cm]{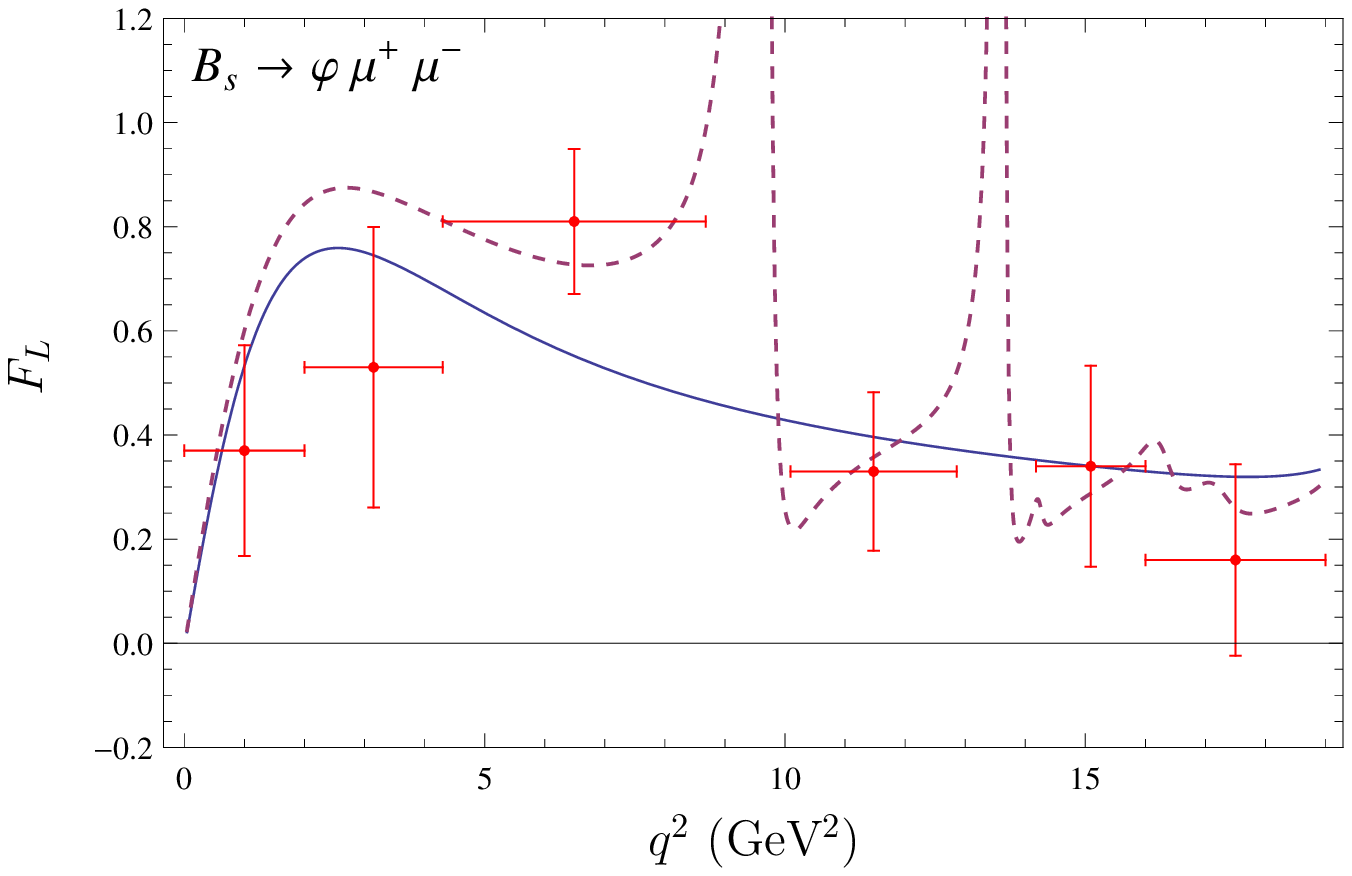}
\caption{Comparison of theoretical predictions for the differential
    branching fractions $d Br(B_s \to \varphi \mu^+\mu^-)/d q^2$ and the $\varphi$ longitudinal polarization
    $F_L$   with available experimental data. }
  \label{fig:brbsphi}
\end{figure}

Now we apply our model for the consideration of the rare $B_s$
decays. Using described above method we explicitly determine the form
factors in the whole accessible kinematical range through the overlap
integrals of the meson wave functions. They again can be approximated
with good accuracy by 
Eqs.~(\ref{fitfv}), (\ref{fita12}). The obtained values of $F(0)$ and
$\sigma_{1,2}$ are given in Table~\ref{hff}. 
\begin{table}
\caption{Calculated form factors of weak $B_s\to \eta_s$ and $B_s\to \varphi$ transitions. }
\label{hff}
\begin{tabular}{ccccccccccc}
\hline
   &\multicolumn{3}{c}{{$B_s\to \eta_s$}}&\multicolumn{7}{c}{{\  $B_s\to \varphi$\
     }}\\
\cline{2-4} \cline{5-11}
& $f_+$ & $f_0$& $f_T$& $V$ & $A_0$ &$A_1$&$A_2$& $T_1$ & $T_2$& $T_3$\\
\hline
$F(0)$          &0.384 &0.384 &  0.301 & 0.406 & 0.322& 0.320 & 0.318 & 0.275& 0.275& 0.133\\
$F(q^2_{\rm max})$&3.31  &0.604 &  1.18 & 2.74& 1.64& 0.652 &  0.980 & 1.47& 0.675& 0.362\\
$\sigma_1$      &$-0.347$&$-0.120$& $-0.897$& $-0.861$ &$-0.104$&  0.133&
1.11  &$-0.491$&$0.396$& $0.639$\\
$\sigma_2$      &$-1.55$&$-0.849$&$-1.34$&$-2.74$&$-1.19$&$-1.02$&
0.105  &$-1.90$&$-0.811$& $-0.531$\\
\hline
\end{tabular}

\end{table}
Using these form factors we consider the rare semileptonic decays. In
the effective Hamiltonian for the $b\to s$ transitions the usual
factorization of short-distance (described by the 
Wilson coefficients) and long-distance contributions (which matrix elements are
proportional to hadronic form factors) is employed. The effective
Wilson coefficient $ c_9^{\rm eff}$ contains additional 
perturbative and long-distance contributions. The long-distance
(nonperturbative) contributions are assumed to originate from the
$c\bar c$ vector resonances ($J/\psi$, $\psi(2S)$, $\psi(3770)$, $\psi(4040)$, $\psi(4160)$ and $\psi(4415)$) and  have the usual
Breit-Wigner structure. In
Fig.~\ref{fig:brbsphi}   we
confront our predictions  for differential branching fractions, $dBr/dq^2$, and the 
longitudinal polarization fraction, $F_L$, with experimental data from
PDG (CDF) \cite{pdg} and recent LHCb \cite{lhcbbr} data.  By solid lines we
show results for the nonresonant branching fractions, where long-distance
contributions  of the charmonium resonances to the
coefficient $c_9^{\rm eff}$ are neglected. Plots given by the dashed lines contain
such resonant contributions. For
decays with the muon pair two largest peaks  correspond to
the contributions coming from the lowest vector charmonium states
$J/\psi$ and $\psi(2S)$, since they are narrow. The region of these resonance
peaks is excluded in experimental studies of these
decays. Contributions in the low recoil region originating from the higher vector
charmonium states, which are above the open charm threshold, are
significantly less pronounced.   The LHCb
values for the differential branching fractions in most $q^2$ bins are  lower
than the CDF ones, but experimental errors are rather large. Our
predictions lie just  in-between these experimental
measurements. For the $\varphi$ longitudinal polarization fraction, $F_L$,
only LHCb data are available which agree with our results within
uncertainties. 

In Table~\ref{brbk} we present our predictions for the nonresonant
branching fractions of the rare semileptonic $B_s$ decays and compare
them with previous calculations \cite{ccf,gl,akf,c,wzz} and available
experimental data \cite{pdg,lhcbbr}. In Ref.~\cite{ccf} the
 form factors were calculated on the basis of the light-cone QCD sum rules within the soft collinear
effective theory. The authors of Ref.~\cite{gl} employ the light front
and constituent quark models for the evaluation of the rare decay
branching fractions. Three-point QCD sum rules are used for the analysis
of the rare semileptonic $B_s$ decays into $\eta(\eta')$ and lepton
pair in Ref.~\cite{akf}. In Ref.~\cite{c} calculations are based on
the light-front quark model, while light-cone sum rules in the
framework of heavy quark effective field theory are applied in
Ref.~\cite{wzz}. The analysis of the predictions given in
Table~\ref{brbk} indicates that  these significantly different
approaches give close values of order  $10^{-7}$ for the rare
semileptonic $B_s\to \varphi(\eta^{(')})l^+l^-$ decay branching
fractions and values of order  $10^{-8}$ for $B_s\to K^{(*)}l^+l^-$ decays.   
Experimental data are available for the branching
fraction of the $B_s\to\varphi\mu^+\mu^-$ decay only. As we see from
the table all 
theoretical predictions are well consistent with each other and
experimental data for the $B_s\to\varphi\mu^+\mu^-$ decay from PDG
\cite{pdg}. Note that very recently the LHCb Collaboration
\cite{lhcbbr} also reported measurement of this decay branching
fraction with the value $7.07^{+0.97}_{-0.94}\times10^{-7}$ which is somewhat lower than previous measurements. Our
prediction is consistent with the latter value within 2$\sigma$.   

\begin{table}
\caption{Comparison of theoretical predictions for the nonresonant
  branching fractions of the rare semileptonic $B_s$ decays and
  available experimental data (in $10^{-7}$). }
\label{brbk}
\begin{tabular}{cccccccc}
\hline
 Decay& this paper & \cite{ccf}& \cite{gl} & \cite{akf}& 
\cite{c}& \cite{wzz}&{Exp.}\cite{pdg} \\
\hline
$B_s\to \eta \mu^+\mu^-$ & $3.8\pm0.4$ &$3.4\pm1.8$&
3.12 & $2.30\pm0.97$ & 2.4 &
$1.2\pm0.12$&\\
$B_s\to \eta \tau^+\tau^-$ & $0.90\pm0.09$&$1.0\pm0.55$& 0.67 & 0.373$\pm$0.156 & 0.58 &
$0.34\pm0.04$&\\
$B_s\to \eta \nu\bar \nu$ & $23.1\pm2.3$&$29\pm15$&
21.7 & $13.5\pm5.6$ & 17 &&\\
$B_s\to \eta' \mu^+\mu^-$ & $3.2\pm0.3$ &$2.8\pm1.5$&
3.42 & $2.24\pm0.94$ & 1.8 &&\\
$B_s\to \eta' \tau^+\tau^-$ & $0.39\pm0.04$&0.47$\pm$0.25& 0.43 & 0.280$\pm$0.118 & 0.26 &&\\
$B_s\to \eta' \nu\bar \nu$ & $19.7\pm2.0$ &$24\pm13$&
23.8 & $13.3\pm5.5$ & 13 & &\\
$B_s\to \varphi \mu^+\mu^-$ & $11.6\pm1.2$ &&16.4 &  &  &
$11.8\pm1.1$& $12.3^{+4.0}_{-3.4}$\\
$B_s\to \varphi \tau^+\tau^-$& $1.5\pm0.2$ &&1.51 &  &  &
$1.23\pm0.11$&\\ 
$B_s\to \varphi \nu\bar \nu$& $79.6\pm8.0$ &&116.5 &  &  & & $<$54000  \\ 
$B_s\to K \mu^+\mu^-$ & $0.24\pm0.03$ && &  &  0.14&
0.199$\pm$0.021&\\
$B_s\to K \tau^+\tau^-$ & 0.059$\pm$0.006 && &  &  0.03&
0.074$\pm$0.007&\\
$B_s\to K \nu\bar\nu$ & $1.42\pm0.14$ && &  &  1.01&&\\
$B_s\to K^* \mu^+\mu^-$ & $0.44\pm0.05$ && &  &  &
$0.38\pm0.03$&\\
$B_s\to K^* \tau^+\tau^-$ & 0.075$\pm$0.008 && &  &  &
0.050$\pm$0.004&\\
$B_s\to K^* \nu\bar\nu$ & $3.0\pm0.3$ && &  &  &&\\
\hline
\end{tabular}

\end{table}

\section{Conclusions}
\label{sec:concl}

The form factors parametrizing the transition matrix elements of the weak current
between the $B_s$ and  heavy ($D_s^{*}$, $D_{sJ}^{(*)}$ ) or light (
$K^{(*)}$,  $K_J^{(*)}$, $\eta(\varphi)$) mesons were
calculated on the basis of the relativistic quark model with the
QCD-motivated quark-antiquark  interaction potential. All relativistic effects,
including boosts of the meson wave functions and contributions of the
intermediate negative-energy states, were consistently taken into
account. The main advantage of the adopted approach consists in 
 that it allows the determination of the momentum transfer
dependence of the form factors in the whole accessible kinematical
range. Therefore no additional assumptions and ad hoc extrapolations
are needed for the description of the weak decays which possess a rather broad kinematical range. This
significantly improves the reliability of the obtained results.

The calculated form factors were used for considering  the
semileptonic and rare $B_s$ decays. The differential
and total decay branching fractions as well as asymmetry and
polarization parameters were evaluated. The obtained results were
confronted with previous investigations based on significantly different
theoretical approaches and available experimental data. An overall good agreement
of our predictions with measured values is observed.

This work was supported in part by the {Russian
Foundation for Basic Research} under Grant No.12-02-00053-a.

\end{document}